\documentclass[review,1p]{elsarticle}
\usepackage{graphics}
\usepackage[ruled]{algorithm2e}
\usepackage{graphicx}
\usepackage{epsfig}

\usepackage{amssymb,amsmath,amsfonts}
\usepackage{amsthm}

\journal{International Journal of Distributed Sensor Networks}

\begin{document}

\begin{frontmatter}



\title{Modelling and Performance analysis of a Network
of Chemical Sensors with Dynamic Collaboration}


\author{Alex Skvortsov}\address{HPP Division, Defence Science and Technology Organisation,
506 Lorimer Street, Fishermans Bend, Vic 3207, Australia }
\author{Branko Ristic}\address{ISR Division, Defence Science and Technology Organisation, 506 Lorimer Street, Fishermans Bend, Vic 3207, Australia}

%
\date{5 January 2010}
\begin{abstract}
The problem of environmental monitoring using a wireless network of
chemical sensors with a limited energy supply is considered. Since
the conventional chemical sensors in active mode consume vast
amounts of energy, an optimisation problem arises in the context of
a balance between the energy consumption and the detection
capabilities of such a network. A protocol based on ``dynamic sensor
collaboration'' is employed: in the absence of any pollutant,
majority of sensors are in the sleep (passive) mode; a sensor is
invoked (activated) by wake-up messages from its neighbors only when
more information is required. The paper proposes a mathematical
model of a network of chemical sensors using this protocol. The
model provides valuable insights into the network behavior and near
optimal capacity design (energy consumption against detection). An
analytical model of the environment, using turbulent mixing to
capture chaotic fluctuations, intermittency and non-homogeneity of
the pollutant distribution, is employed in the study. A binary model
of a chemical sensor is assumed (a device with threshold detection).
The outcome of the study is a set of simple analytical tools for
sensor network design, optimisation, and performance analysis.
\end{abstract}

\end{frontmatter}

\renewcommand{\baselinestretch}{1.5}

\section{Introduction}

Development of wireless sensor network (WSN) for a particular
operation scenario is a complex scientific and technical problem
\cite{R0}, \cite{R14}. Very often this complexity resides in establishing a
balance between the peak performances of the WSN prescribed by the
operational requirements (e.g. minimal detection threshold, size of
surveillance region, detection time, rate of false negatives, etc)
and various resource constrains (e.g. limited energy supply, limited
number of sensors, limited communication range, fixed detection
threshold of individual sensors, limited budget for the cost of
hardware, maintenance, etc). The issue of resource constraints
becomes even more relevant for a network of chemical sensors that are used for the continuous environmental  monitoring (air and water pollution, hazardous releases, smoke etc).  The
reason is that a modern chemical sensor is usually equipped with a
sampling unit (a fan for air and a pump for water), which turns on when the sensor is active.
The sampling unit usually requires a significant amount of
energy to operate as well as frequent replacement of some consumable
items (i.e. cartridges, filters). This leads to the critical
requirement in the design of a WSN to reduce the active
(i.e. sampling) time of its individual sensors.

One attractive way to achieve an optimal balance between the peak
performance of the WSN and its constraints in resources, mentioned
above is to exploit the idea of Dynamic Sensor Collaboration (DSC)
\cite{R1}, \cite{R2}. The DSC implies that a sensor in the network
should be invoked (or activated) only when the network will gain
information by its activation \cite{R2}. For each individual sensor
this information gain can be evaluated against other performance
criteria of  the sensor system, such as the detection delay or
detection threshold, to find an optimal solution in the given
circumstances.

While the DSC-based approach is a convenient framework for the
development of algorithms for optimal scheduling of constrained
sensing resources, the DSC-based algorithms involve continuous
estimation of the state of each sensor in the network and usually
require extensive computer simulations \cite{R1}, \cite{R2}. These
simulations may become unpractical as the number of sensors in the
network increases (e.g. ``smart dust'' sensors). Even when feasible,
the simulations can provide only the numerical values for optimal
network parameters, which are specific for an analysed scenario, but
without any analytical framework for their consistent interpretation
and generalisation. For instance, the scaling properties of a
network (the functional relationship between the network parameters)
still remain undetermined, which prevents any comprehensive
optimisation study.

This motivates the development of another, perhaps less rigorous,
but certainly simpler approach to the problem of network analysis
and design. The main idea is to phenomenologically employ the
so-called bio-inspired (epidemiology, population dynamics) or
physics inspired (percolation and graph theory) models of DSC in the
sensor network in order to describe the dynamics of collaboration as
a single entity  \cite{R22}, \cite{R3}, \cite{R5}, \cite{R4},
\cite{R6},\cite{R7}. Since the theoretical framework for the bio- or
physics- inspired models is already well established, we are in the
position to make significant progress in the analytical treatment of
these models of DSC (including  their optimisation). From a formal
point of view the derived equations are ones of the ``mean-field''
theory, meaning that instead of working with dynamic equations for
each individual sensor  we only have a small number of equations for
the ``averaged'' sensor state (i.e. passive, active, faulty etc),
{\em regardless of the number of the sensors in the system}. A
reveling example of the efficiency of this approach is the
celebrated SIR model in epidemiology \cite{R9}. For any size of
population, the SIR model describes the spread of an infection by
using only three equations, corresponding to three ``infectious''
classes  of the population: susceptible, infectious and recovered.

The analytic or ``equation-based'' approach, often leads to valuable
insights into the performance of the proposed sensor network system
by providing simple analytical expressions to calculate the vital
network parameters, such as detection threshold, robustness,
responsiveness and stability and their functional relationships.

In the current paper we develop a simple model of a wireless network
of chemical sensors, where dynamic sensor collaboration is driven by
the level of concentration of a pollutant (referred to as the
``external challenge'') at each individual sensor. Our approach is
based on the known analogy \cite{R7} between the information spread
in a sensor network and the epidemics propagation across a
population. In this analogy, the infection transmission process
corresponds to message passing among the sensors. A chain reaction
in transmission of an infection is called the epidemic. In the
context of a sensor network, a chain reaction will trigger the
network (as a whole) to move from the ``no pollutant'' state to the
``pollutant present'' state, which will indicate the presence of an
external challenge.

The paper shows that the adopted epidemics or population inspired
approach can provide a reliable description of the dynamics of such
a sensor network. The simple analytical formulas (scaling laws)
derived from the model express the relationships between the
parameters of the network (e.g. number of sensors, their density,
sensing time etc), the network performance (probability of
detection, response time of a network) and the parameters of the
external challenge (environment, pollutant). As an example of
application of the proposed framework we performed a simple
optimisation study. Numerical simulations are carried out and
presented in the paper in support of analytic expressions.

Although the model presented in this paper is specific to a network
of chemical sensors, the underlying analytical approach can be
easily adapted to other applications and other types of networks by
a simple change of the model of environment and sensor.

\section{The Model of Environment}

%
%

The external challenges are modeled by a random time series which
mimics the turbulent fluctuation of concentration at each sensor of
the network. In this approach the fluctuations in concentration $C$
are modeled by the probability density function (pdf) of $C$ with
the mean ${C_0}$ as a parameter (i.e. ${C_0}$ is a mean
concentration of the tracer in the area) \cite{bisiganesi_07} :

\begin{equation}\label{eqn:E0}
    f(C|C_0)  =  (1-\omega)\delta(C) +
    \frac{\omega^2}{C_0}\frac{(\gamma-1)}{(\gamma-2 )}\left(
    1 +
    \frac{\omega}{(\gamma-2)}\frac{C}{C_0}\right)^{-\gamma}.
\end{equation}
Here the value $\gamma = 26/3$ can be chosen to make it compliant
with the theory of tracer dispersion in Kolmogorov turbulence (see
\cite{bisiganesi_07}), but it may vary with the meteorological
conditions. The parameter $\omega$, which models the tracer
intermittency in the turbulent flow, can be in the range $[0,1]$,
with $\omega =1$ corresponding to the non-intermittent case. In
general it also depends on a sensor position within a chemical
plume, thus $\omega$ is in the range $0.95 - 0.98$ near the plume
centroid and  may drop to $0.3 - 0.5$ near the plume edge. For
$\omega \neq 0$, the pdf $f$ of (\ref{eqn:E0}) has a delta impulse
in zero, meaning that the measured concentration in the presence of
intermittency can be zero on some occasions. It can be easily shown
that the pdf of (\ref{eqn:E0}) integrates to unity, so it is
appropriately normalized.

\begin{figure}[h] \label{F:F1}
  {\includegraphics[width=3.5in,height=1.2in]{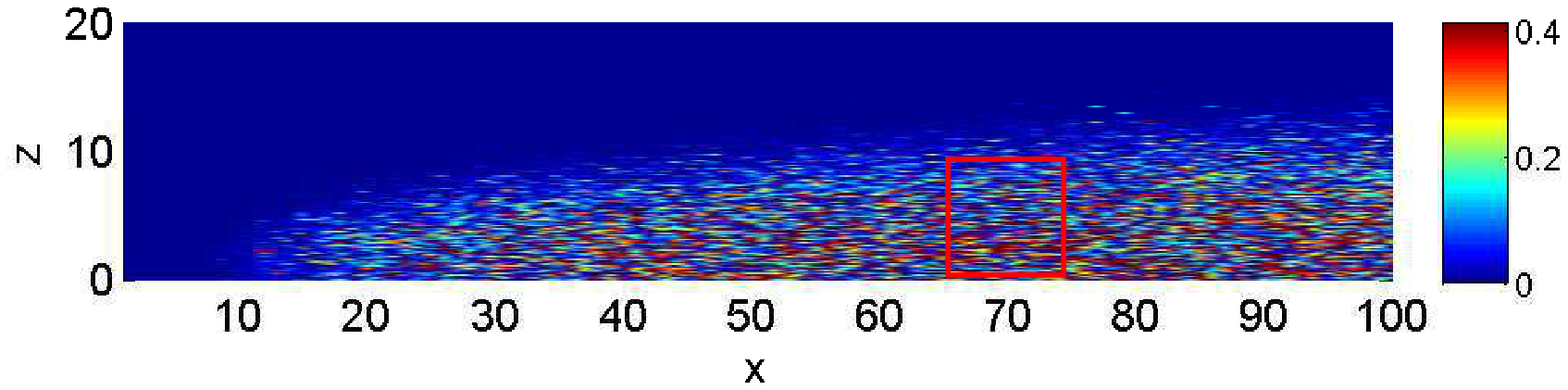}}
  {\includegraphics[width=1.3in,height=1.5in]{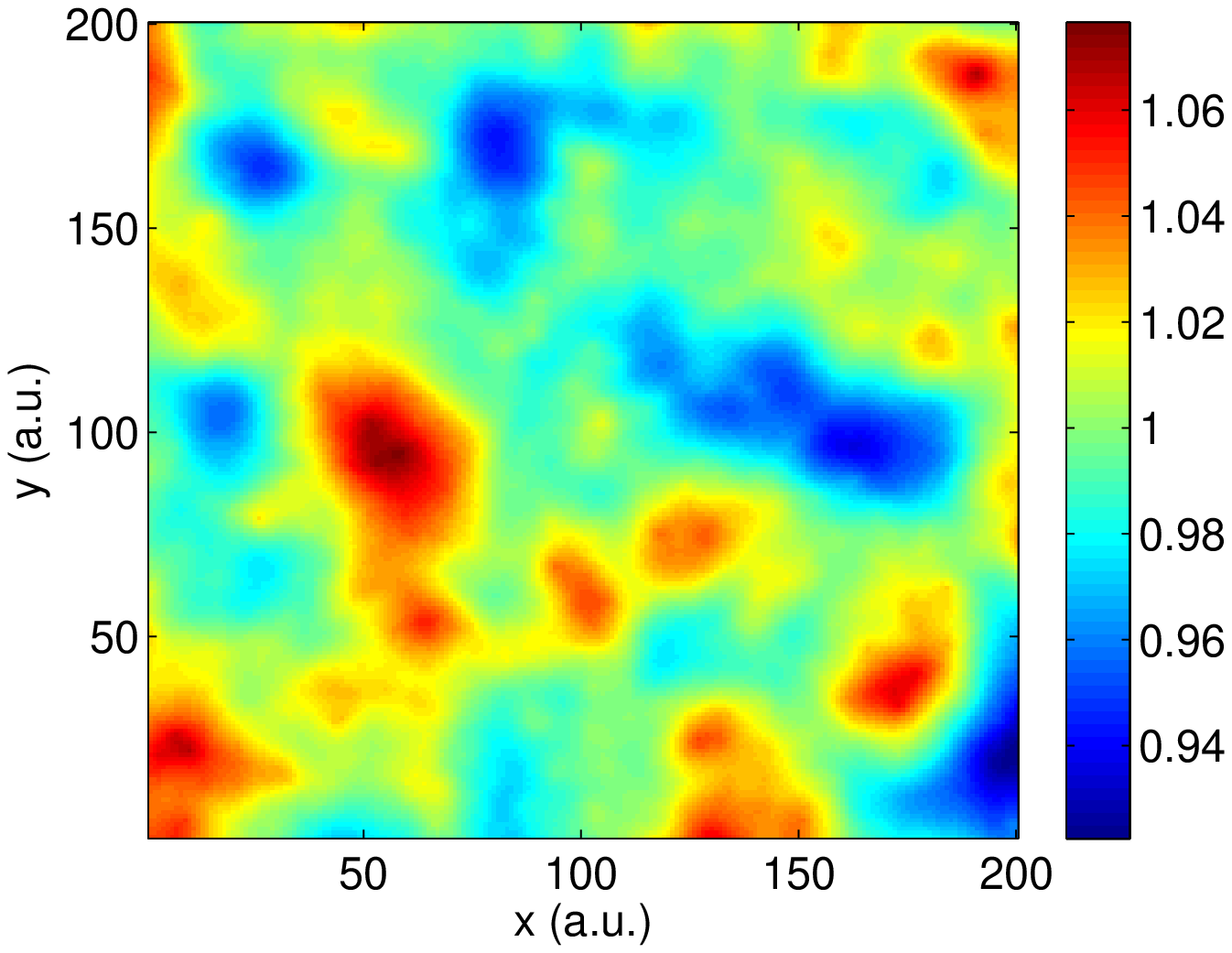}}
    \caption{Example for concentration realisation for a plume-like
    flow (left) and the selected area of the flow with higher resolution (right).}
\end{figure}

The measured concentration time series can be generated by drawing
random samples from the probability density function given in
(\ref{eqn:E0}) at each time step. The random number generator is
implemented using the \emph{inverse transform} method
 based on the following steps \cite{R00} :
\begin{enumerate}
  \item Draw a sample $u$ from the standard uniform distribution:
  $u\sim U[0,1]$.
  \item Compute the value of $C$ that satisfies $F(C)=u$, where
  $F(\cdot)$ is the cumulative distribution function (cdf) of the
  distribution of interest.
  \item The value of $C$ computed in the previous step is a random sample drawn from the desired
  probability distribution.
\end{enumerate}

The cdf $F(\cdot)$ needed for inverse transform sampling is obtained
by integrating the pdf in (\ref{eqn:E0}), and is given by:
\begin{equation}\label{eqn:E1}
    F(C|C_0) =
    1-\omega \left[1+\left(\frac{2}{\gamma-2}\right)\frac{C}{C_0}\right]^{1-\gamma}.
\end{equation}
The use of this cdf in the inverse transform sampling procedure
generates the value of concentration:
\begin{equation}\label{eqn:E2}
  C = \begin{cases} C_0 \left(\frac{\gamma-2}{\omega}\right)
  \left[ \left( \frac{1-u}{\omega}\right)^{-\frac{1}{\gamma-1}}-1\right], &u \geq 1-\omega \\
   0, &u < 1-\omega,\end{cases}
\end{equation}
where  $u$ is again the standard uniform distribution $u\sim
U[0,1]$.

In order to produce spatial correlations that compline with the well-known scaling properties of turbulent dispersion
a special `swapping' algorithms was implement. This recursive algorithm mimic the chaotic fluctuations occurring in the real turbulent flows (for details, see \cite{R11}).

The proposed framework allows to implement  a reasonably realistic
model of the contaminated environment (i.e to generate the
concentration realisation at each sensor over time), see Fig 1. Due
to a universal nature of turbulence it can be used to simulate
performance of WSN in detection of either airborne and  waterborne
releases. The parameters $\gamma$ and $\omega$ are typically
estimated from geophysical observation (meteorological and
organological) and will be assumed known.

The geometrical complexity of the turbulent flow can be incorporated in the theoretical framework  (\ref{eqn:E1}) by assuming a temporal and spatial variability of the mean concentration filed $C_0 \equiv C_0 (\textbf{r}, t)$. This way we can simulate various morphologies of the flow (jet, wake, boundary layer, compartment flow, etc) as well as various scenarios of hazardous release (plume, puff ), for details see \cite{R00} , \cite{R12}. For the sake of simplicity in the current paper we consider only case $C_0 = const$. This assumption corresponds to the approximation when the size of WSN is less that the width of hazardous plume (see Fig 1), or to an important practical case of a `highly distributed'  source of pollutant (traffic, extended industrial site or urban area \cite{R13}).


\section{The Model of a Chemical Sensor}

We adopt a simple binary (or ``threshold'')  model of a sensor, with
the sensor reading $V$ given by:
\begin{equation}\label{eqn:E3}
  V = \begin{cases} 1, & C \geq C_{*} \\
   0, & C < C_{*}. \end{cases}
\end{equation}
We emphasize that threshold $C_{*}$ is an internal characteristic of
the sensor, unrelated to $C_0$ in (\ref{eqn:E0}). This threshold is
another important parameter of our model. A chemical sensor with bar
readings, which includes many subsequent levels for concentration
thresholds mapped into a discrete sensor output, is an evident
generalisation of (\ref{eqn:E3}).

Using (\ref{eqn:E2}) and (\ref{eqn:E3})  it is straightforward to
derive the probability of detection for an individual sensor
embedded in the environment characterised by (\ref{eqn:E1}):
\begin{equation}\label{eqn:E4}
  p = 1 - F (C_{*}|C_0).
\end{equation}
This aggregated parameter links the characteristics of a specific
sensor $C_{*}$, the parameter of the external challenge $C_0$ and
the environment ($F (\cdot), \gamma, \omega$).

\section{Modeling and Analysis of Network Performance}

Our focus is a wireless network of chemical sensors with dynamic
collaboration. We assume that $N$ identical sensors (i.e with the same detection threshold $C_*$ and sampling time $\tau_*$) are uniformly distributed
over the surveillance domain of area $S$ with density $\rho = N/S$.


 We will model the following network protocol for dynamic collaboration.
Each sensor can be only in one of the two states: {\em active} or
{\em passive}. The sensor can be activated only by a message it
receives from another sensor. Once activated, the sensor remains in
the active state during an interval of time $\tau_{*}$; then it
returns to the passive (sleep) state. While being in the active
state, the sensor senses the environment and if the chemical tracer
is detected (binary detection), it broadcasts a (single) message. If
a sensor receives an activation message while it is in the active
state, it will ignore this message. The broadcast capability of the
sensor is characterized by its communication range $r_*$, which is
another important parameter of the model. The described protocol
assumes that certain sensors of the network are permanently active.
The number of permanently active sensors in the network is fixed but
the actual permanently active sensors vary over time in order to
equally distribute the energy consumption of individual sensors.

%

The  WSN following this protocol  can be considered as a system of
agents, interacting with each other (by means of message exchange)
and with the stochastic environment (by means of sampling and
probing). The interactions can change the state of agents (active
and passive). From this perspective this  WSN  is similar to the
epidemic SIS (susceptible-infected-susceptible) model \cite{R9}, in
which an individual can be in only two states (susceptible or
infected), and the change of state is a result of interaction
(mixing) between the individuals (which corresponds to the exchange
of messages in our case). Thus a dynamic (population) model for our
system \cite{R9} is as follows:
\begin{eqnarray}\label{eqn:E5}
      \frac{d N_{+}}{dt} & = &  \alpha N_{+} N_{-}  -
      \frac{N_{+}}{\tau_*}, \\
      \frac{d N_{-}}{dt} & = &  - \alpha N_{+} N_{-}  +
      \frac{N_{+}}{\tau_*}, \label{eqn:E5a}
\end{eqnarray}
where $N_{+}, N_{-}$ denote the number of active and passive
sensors, respectively. The nonlinear terms on the RHS of
(\ref{eqn:E5}) and (\ref{eqn:E5a}) are responsible for the
interaction between individuals (i.e. sensors), with the parameter
$\alpha$ being a measure of this interaction. The population size
(i.e. the number of sensors) is conserved, that is $N_{+} + N_{-} =
N = const$.

The next step is to express $\alpha$ in terms of the parameters of
our system by invoking physics based arguments used in population dynamics \cite{R9} . It is well-known that parameter $\alpha$ in  (\ref{eqn:E5}) describes the intensity (contact rate) of social interaction between individuals in the community, so we can propose (see \cite{R9}, \cite{R10})
\begin{equation}
\label{eqn:E110}
      \alpha \propto   \frac{m p }{N \tau_{*}},
\end{equation}
where $m$ is the number of contacts  made by an ``infected'' sensor
during the infectious period $\tau_*$ (i.e the number of sensors
receiving a message from an alerting sensor). In our case we have $m
= \pi r^2_{*} \rho$. Then using  $N = S \rho$ we can write
\begin{equation}
\label{eqn:E66}
      \alpha  = G \frac{\pi r^2_{*}}{\tau_* S} p ,
\end{equation}
where $G$ is a constant calibration factor, being of order unity (it
must be estimated during the network calibration); $p$ was defined
by (\ref{eqn:E4}). In order to simplify notation, from now on  we
will assume that $G$ is absorbed in the definition of $r_*$.

It is worth noting that by introducing non-dimensional variables $n_+ = N_+ /N, n_- = N_- /N, \tau = t/\tau_{*} $ the system (\ref{eqn:E5})-(\ref{eqn:E5a}) can be rewritten in a compact non-dimensional form
\begin{eqnarray}
\label{eqn:E661}
      \frac{d n_{+}}{d \tau } & = &  R_0 n_{+} n_{-}  - n_{+}, ~~ n_{-} = 1 - n_{+},
\end{eqnarray}
with only one non-dimensional parameter
\begin{eqnarray}
\label{eqn:E662}
     R_0 = \alpha \tau_{*}  N.
\end{eqnarray}
The parameter $R_0$ is well-known in epidemiology where it
has the meaning of a {\em basic reproductive number} \cite{R9}.

The system (\ref{eqn:E5})-(\ref{eqn:E5a}) combined with the
condition $N_{+} + N_{-} = N$ can be reduced to one equation for $y
= N_{+}$
\begin{equation}\label{eqn:E104}
      \frac{d y}{dt} =    \alpha y (N - y)  - \frac{y}{\tau_*} = y (b
      - \alpha y),
\end{equation}
where
\begin{equation}\label{eqn:E1041}
      b = \alpha N - 1/\tau_* = (R_0 -1)/\tau_*.
\end{equation}
By simple change of variables $z = \alpha y/b$ this equation can be
reduced to the standard logistic equation
\begin{equation}\label{eqn:E105}
      \frac{d z}{dt} =  b z (1-z),
\end{equation}
which has the well-known solution
\begin{equation}\label{eqn:E106}
      z (t) = \frac{z_0}{(1 - z_0)\exp (- b t) + z_0},
\end{equation}
where $z_0 =  z(0)$.

We can see that if $b < 0$ then $z \rightarrow 0$ as $t \rightarrow
\infty$ for any $z_0$, so any individual sensor activation in the
network will ``die out'', that is the network will not be able to
detect the external challenge. The same is valid for $b = 0$ when $z
= z_0 = const$ (no response to external challenges). Only if the
condition  $b  > 0$ is satisfied, then $z \rightarrow 1$ as $t
\rightarrow \infty$ (independently of $z_0$). In this case, after a
certain transition interval, the network will reach a new steady
state with
\begin{equation}\label{eqn:E1061}
     \frac{N_{+}}{N} = 1 - \theta, ~~    \frac{N_{-}}{N} = \theta,
     ~~ \theta  = \frac{1}{ \alpha \tau_{*}  N} \equiv   \frac{1}{R_0}.
\end{equation}

A fraction of active sensors $N_{+}$ at this new state is a measure
of the network (positive) response to the event of chemical
contamination. From (\ref{eqn:E106}) it is clear that the time scale
for the network to reach the new state can be estimated from the condition $ e^{- b t} \ll 1 $, so
\begin{equation}\label{eqn:E1062}
     \tau  \geq \frac{1}{b} = \frac{\tau_{*}}{R_0 - 1}.
\end{equation}
This equation provides the relationship between the  scale of activation time and  parameter $R_0$. One can see that this scale decreases as $R_0$ increases.

From (\ref{eqn:E105}), (\ref{eqn:E1062}) it follows that an ``epidemic threshold'' for
the sensor network is simply $\alpha \tau_{*} N  > 1$ or in terms of the 'basic reproductive number'  (\ref{eqn:E662})
\begin{equation}
\label{eqn:E1063}
       R_0 = \alpha \tau_{*}  N  =   p N  \frac{\pi r^2_{*}}{S}  >
       1.
\end{equation}
Observe that sensor sampling time $\tau_{*}$ has disappeared from
the expression for $R_0$. This means that it is possible to create
an information epidemic (i.e. detect a chemical pollutant) for any
value of $\tau_{*}$, provided this time is long enough for a sensor
to detect the chemical tracer. But according to  (\ref{eqn:E1062}),
the responsiveness of the whole network to the external challenges
(i.e. the time constant of detection) is, indeed, strongly dependent
on the sensor sampling time $\tau = \tau_{*}/(R_0 - 1)$.

The expressions (\ref{eqn:E1061}), (\ref{eqn:E1062}) and
(\ref{eqn:E1063}) are the main analytical results of the paper. For
a given level of external challenges (i.e. $C_0$) and meteorological
conditions (i.e. $\gamma,\omega$), these expressions provide a
simple yet rigorous way to estimate how a change in the network and
sensor parameters (i.e. $N, C_{*}, \tau_{*}$) will affect the
network performance (i.e. {$N_{+}, \tau$}). We can also see that for
a given external challenge the network of chemical sensors will
respond in the most effective way when its parameters are selected
in the combination which meets the criterion for `information
epidemic'' (\ref{eqn:E1063}).

The final analytical expressions enable us to maximize the network
information gain and optimize other parameters. For example, from
(\ref{eqn:E1061}), we can readily infer the important scaling
properties of the network performance:
\begin{equation}
 \label{eqn:E118}
   \frac{N_{-}}{N} \sim \frac{1}{r_{*}^2}, ~~ \frac{N_{-}}{N} \sim \frac{1}{N}, ~~ \frac{N_{-}}{N} \sim
   \frac{1}{p}.
\end{equation}
For instance, if we double the communication range of an individual
sensor $r_*$, the fraction of inactive sensors in the network will
drop four times. Likewise, if we need to reach a specified fraction
of active sensors ($1 - N_{-}/N$) to be able to reliably detect a
given level of pollutant concentration, these formulas describe all
possible ways of changing the parameters of the model in order to
achieve this goal.

\section{Information Gain of Collaboration}

We have explained earlier that the concept of DSC is important for a
network with limited energy/material resources.  But the question
remains will a network with DSC be inferior (in terms of detection
performance) in comparison with a benchmark network where all
sensors operate independently of each other and only report their
(positive) detections of chemical pollution to the central processor
for decision making? Clearly, such a benchmark network would be very
expensive to run (all sensors would have to be active all the time),
but could provide excellent detection performance.

In this section we show that, under a certain condition, the network
with DSC can provide superior detection performance compared to the
benchmark network. Let us assume that we have $\delta N$ senors
continuously operating ($0\leq \delta \leq 1 $).  For a benchmark
network, on average, we have  $p \delta N$ sensors detecting
pollutant. For the  network with DSC the same quantity can be
estimated as $p(1-\theta)N$ (since as we have seen the saturation
level of $N_{+}$ does not depend on initial conditions). From here
we can then deduce that the network with DSC will provide more
information (for detection of chemical pollution) than the benchmark
network if the following condition is satisfied:
\begin{equation}
\label{eqn:E1181}
 \theta = \frac{1}{\alpha \tau_{*} N} \leq  (1 - \delta),
 \end{equation}
which is eventually reduced to the condition of ``epidemic
threshold''(\ref{eqn:E1063}) for the small value of $\delta$.

The value of the parameter $\delta$ can be also estimated based on
the following arguments. Let us assume that our aim is to detect  a
level concentration $C_0$ associated with a hazardous release within
the time $T$ (the constraint on time is driven by the requirement to
mitigate the toxic effect of the release). Then we can write a
simple condition for the information `epidemic' in the WSN to occur
during time $T$:
\begin{equation}
\label{eqn:E1182}
 \delta p N T/\tau_* \geq 1,
\end{equation}
where $p$ is given by (\ref{eqn:E4}), i.e. $p = 1 - F (C_{*}|C_0)$.
Evidently,  for information epidemic to be observable, the number of
continuously active sensors should be less that the number of
sensors activated due to the hazardous release. Thus from
(\ref{eqn:E1181}) we can write the following `consistency' condition
for the minimum value of $\delta$
\begin{equation}
\label{eqn:E1182a}
 \delta_{\min}  \approx   \frac{\tau_*}{p N  T } \le   (1 - \frac{1}{\alpha \tau_{*} N} ) ,
\end{equation}
or by re-writing it in terms of $R_0$, see (\ref{eqn:E1061}),
\begin{equation}
\label{eqn:E1183}
   \delta_{\min}  \approx  \frac{\tau_*}{p N T }  \le (1 -   \frac{1}{R_0} ) .
\end{equation}
It can be seen, that  with other conditions being equal  the
fraction of `stand-by' sensors  $\delta_{\min}$ can be made however
small (since $R_0 \geq 1$). It implies that only a small fraction of
WSN will be active most of the time and is a clear demonstration of
the energy consumption gain associated with the `epidemic' protocol.

Another important criteria for epidemic protocol can be derived by comparison of amplitude
of ``detectable events'' for the \emph{same number of sensors} in
the network with DSC with the system of $N$ independent sensors. For
the network with DSC it is $(1-\theta)N$ (since we use $N_{+}$ to
retrieve information about the environment) and for the system of
the \emph{same independent sensors} it is still $pN$ (since $N_{+}$
is simply equal to $N$). Then instead of (\ref{eqn:E1181}) we can
write
\begin{equation}
 \theta < (1 - p)
 \end{equation}
Under this condition more detectable events will occur in the
presence of chemical pollution by the described network with DSC
(activation messages) then in a network of stand alone sensors
(signals of positive detection). This leads to the interesting
threshold condition on the number of sensors in the network
\begin{equation}
 \label{eqn:E1141}
  N > \frac {S}{\pi r_{*}^2} \frac{1}{p (1-p)}.
 \end{equation}
The last term  in RHS $(p (1-p))^{-1}$ has an obvious minimum $4$
corresponding to $p =1/2$, so finally we arrive at the simple
universal condition
\begin{equation}
 \label{eqn:E115}
  N > N_{*} = \frac{4}{\pi}\frac {S}{r_{*}^2}.
 \end{equation}
This condition reads that if the number of sensors in the system is
greater than $N_{*}$ then networking with DSC \emph{can} provide an
information gain over the benchmark network. Under this condition,
the network with DSC in not only desirable from the aspect of energy
conservation, but also provides better detection performance through
the information gain.

The condition $p =1/2$ minimizing RHS of (\ref{eqn:E1141}) can be
considered as a criterion for an ``optimal'' sensor for a given
network with DSC and  for a given concentration of pollutant to be
detected. Namely, from the equation $F (C_{*}|C_0) =1/2$ and using
(\ref{eqn:E1}) we can write
\begin{equation}
 \label{eqn:E117}
   C_{*} = C_0 \left(\frac{\gamma  - 2}{2}  \right) \left[ \left( \frac{1}{2 \omega}\right)^{1 / (1 - \gamma ) }  -
   1 \right].
 \end{equation}
Given environmental parameters ($\gamma,\omega$) and given the level
of concentration to be detected ($C_0$), formula (\ref{eqn:E117})
also specifies a simple condition on detection threshold for an
individual sensor to maximize an information gain by being
networked.

\section{Numerical Simulations}

In support of analytical derivations presented above, a network of
chemical sensors operating according to the adopted protocol for
dynamic collaboration was implemented in MATLAB.    A comprehensive
report with numerical simulations result will be published
elsewhere; here we present only some illustrative examples.

For consistency, a $1000\mbox{m} \times 1000$m surveillance region
populated by $N=400$ sensors with a uniformly random placement was
assumed in all tests. In each run, chemical pollution with
concentration $C_0=150$ is applied, and the simulation starts when a
single randomly selected sensor (which has detected the presence of
chemical contamination in its vicinity) starts broadcasting. Due to
this random initiation and the fact that the probability of
detection of individual sensors is less than unity ($p<1$), each run
of the computer program results in a slightly different outcome.
Figs.\ref{F:F1},\ref{F:F11}  show the average evolution of the ratio
$N_+/N$ in the network over time. The curves were obtained by
using the following parameters:
$\omega=0.98$, $\gamma=26/3$. Fig.\ref{F:F1} demonstrates the
changes in dynamics of the WSN for different values of communication
range $r_*$ and  Fig.\ref{F:F11} depicts the similar plots for
changes of the detection threshold of individual sensor $C_*$. For
all plots in Fig.\ref{F:F1}, Fig.\ref{F:F11} the initial number of active sensors $N_+ (t=0) =10$.

Overall we found that the simulation output is much more sensitive
to the changes of communication range, than to the threshold of an
individual sensor (see range of parameters depicted in
Figs.\ref{F:F1},\ref{F:F11}). In all cases we observed the
transition of $N_+$ from the initial steady state (where $N_+$ is
very small indicating the absence of the pollutant) to the new
steady state (high value of $N_+$), so information ``epidemic''  in
the network of chemical sensors does occur. By direct substitution
into (\ref{eqn:E1063}) it was also validated that in all cases
presented in Figs.\ref{F:F1},\ref{F:F11} the condition for an
information ``epidemic'' was satisfied. In general, the saturation
value of $N_+$ derived from these plots were in an agreement with
theoretical prediction (\ref{eqn:E1061}), but the estimated
standard deviation of $N_+$ (not shown in Fig.\ref{F:F1}) could be
very high (up to $30 \%$) for some combination of parameters. The
relative standard deviation (normalized by mean value $N_+$) usually
gradually decreased over time and quite rapidly decays with the
increase of communication range $r_*$. The occasional high
variability of the output of the sensor network is undesirable and
motivates further analysis. We also used the data from the plots in
Figs. \ref{F:F1}, \ref{F:F11} to calibrate our model. The
calibration was performed by extracting the  steady-sate (or
saturation) values of $N_+$ from the plots and by adjusting the
``free'' constant $G$ in the analytical expressions
(\ref{eqn:E1061}) to achieve the best match between the analytical
predictions and simulations. The value $G\approx 0.7$ seems to
provide an optimal agreement with the presented simulations.

In order to validate our simple model for parameter $\alpha$ we
performed the following study. For each simulation we derived value
of $\alpha_s$  from (\ref{eqn:E1061})   and  then compared it with
the value $\alpha_t$ calculated from the theoretical expression
(\ref{eqn:E66}) using the  calibration value $G\approx 0.7$. The
results of this study are presented in Fig.\ref{F:F111}. The red
dashed line corresponds to the perfect agreement between the theory
and simulations. Considering the high variability of $N_+$ and a
rather simple model for $\alpha$, the agreement between the theory
and simulations is acceptable.

\begin{figure}
  \centerline{\includegraphics[width=4.2in]{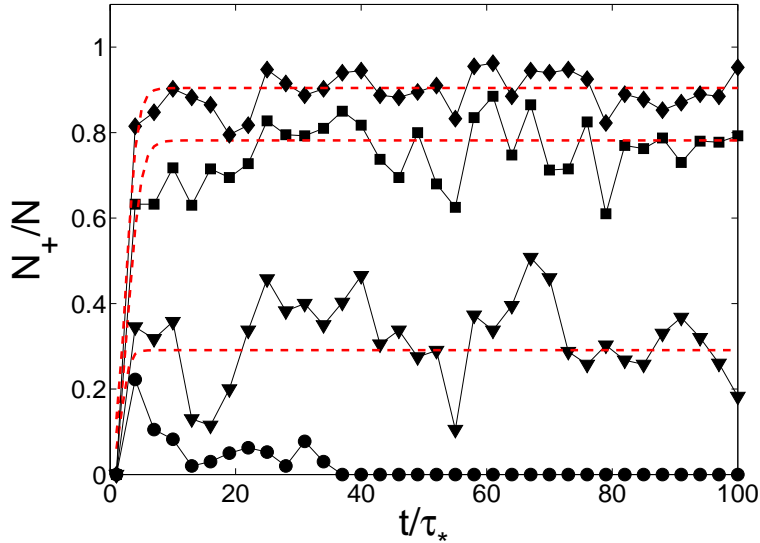}}
    \caption{Results of numerical simulations: The fraction of active sensors in the network over time for the different communication range $r_*$: $\diamondsuit - r_*=40m $,$~\Box- r_*=30m $, $~\bigtriangledown - r_*=27m $, $~\bigcirc - r_*=20m$; $C_*/C_0 = 1.03$, $N_+ (t =0) = 10$. The dashed red line corresponds to the analytical predictions (\ref{eqn:E106}). It is clearly seen that in the case $r_*=20m$ the information epidemic in WSN dies off.}
\label{F:F1a}
\end{figure}

\begin{figure}
  \centerline{\includegraphics[width=4.2in]{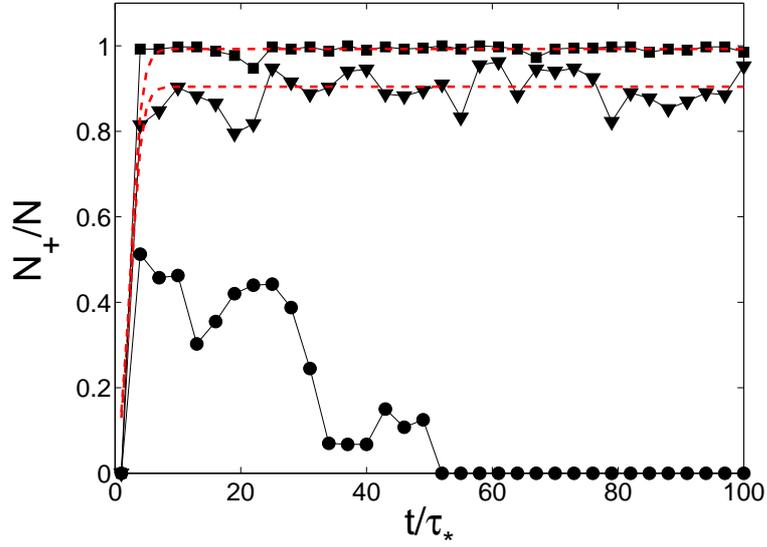}}
   \caption{Results of numerical simulations: The fraction of active sensors in the network over time for the different threshold of individual sensor  $C_*$: $~\Box - C_*/C_0 = 1.05$, $~\bigtriangledown - C_*/C_0 = 1.02$, $~\bigcirc - C_*/C_0 = 1.00$; $ r_*=40m$, $N_+ (t =0) = 10$. The dashed red line corresponds to the analytical predictions (\ref{eqn:E106}). It is clearly seen that in the case $C_*/C_0 = 1.00$ the information epidemic in WSN dies off.}
\label{F:F11}
\end{figure}

\begin{figure}
  \centerline{\includegraphics[width=3.2in]{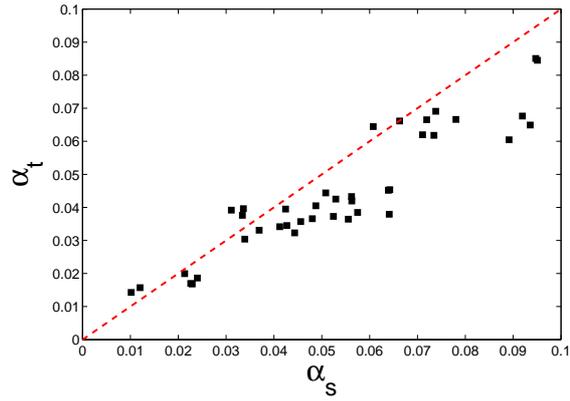}}
    \caption{Simulation and theoretical predictions of parameter $\alpha$ : $\alpha_t$ is the theoretical value (\ref{eqn:E66}), $\alpha_s$  is the results of simulations.  The red dashed line corresponds to the perfect agreement.}
\label{F:F111}
\end{figure}

\begin{figure}
  \centerline{\includegraphics[width=3in]{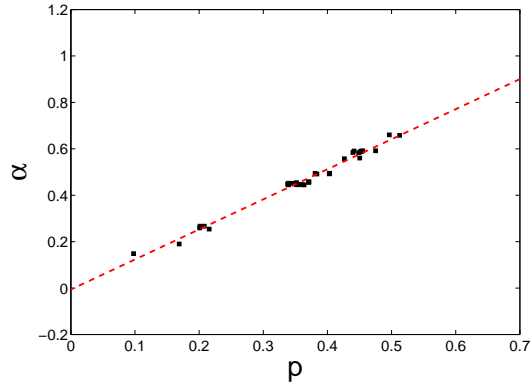}}
    \caption{Parameter $\alpha$ as a function of $p$ extracted from numerical simulations in log-log scale.
    The dashed line corresponds to the power-law fit $\alpha \propto p^{q}, q =1.27$, theoretical prediction corresponds to
     $q =1$, see (\ref{eqn:E110}). }
\label{F:F2}
\end{figure}

To validate further the alignment between the computer simulations
and the proposed mathematical model, we  numerically estimated some
scaling properties of the network system (i.e. (\ref{eqn:E66}),
(\ref{eqn:E118})). Firstly we derived the scaling properties from
computer simulations and then compared them to the theoretical
predictions. In general we found that all trends of the derived
scaling do agree with theoretical expressions in (\ref{eqn:E118}),
but the quantitative agrement may significantly vary from case to
case.  As an illustration, in Fig.\ref{F:F2} we present the plot of
dependency of $\alpha$ against $p$ in log-log scale. The extracted
exponent corresponds to $\alpha \propto p^q$, where $q = 1.27$ while
the theoretical value according to (\ref{eqn:E110}) is $q = 1$. This
indicates that while our analytical model is very simple and fast to
compute, for higher accuracy it may need further refinements as
discussed below.

The results of numerical simulations presented above   serve to
verify that the `information epidemic' does occur in the wireless
network of chemical senors. This also implies that the proposed
theoretical framework may lead to a gain in the energy consumption,
that may result in the significant advantages in operational
deployment of such systems. More detailed analysis of the optimal
values of parameters satisfying threshold conditions
(\ref{eqn:E1063}), (\ref{eqn:E1183}), (\ref{eqn:E115})  and  lead to
the optimal performance of WSN will be reported in separate
publications.

\section{Refinements of the model}

\begin{figure}[h]
  \centerline{
  {\includegraphics[width=2.2in,height=1.6in]{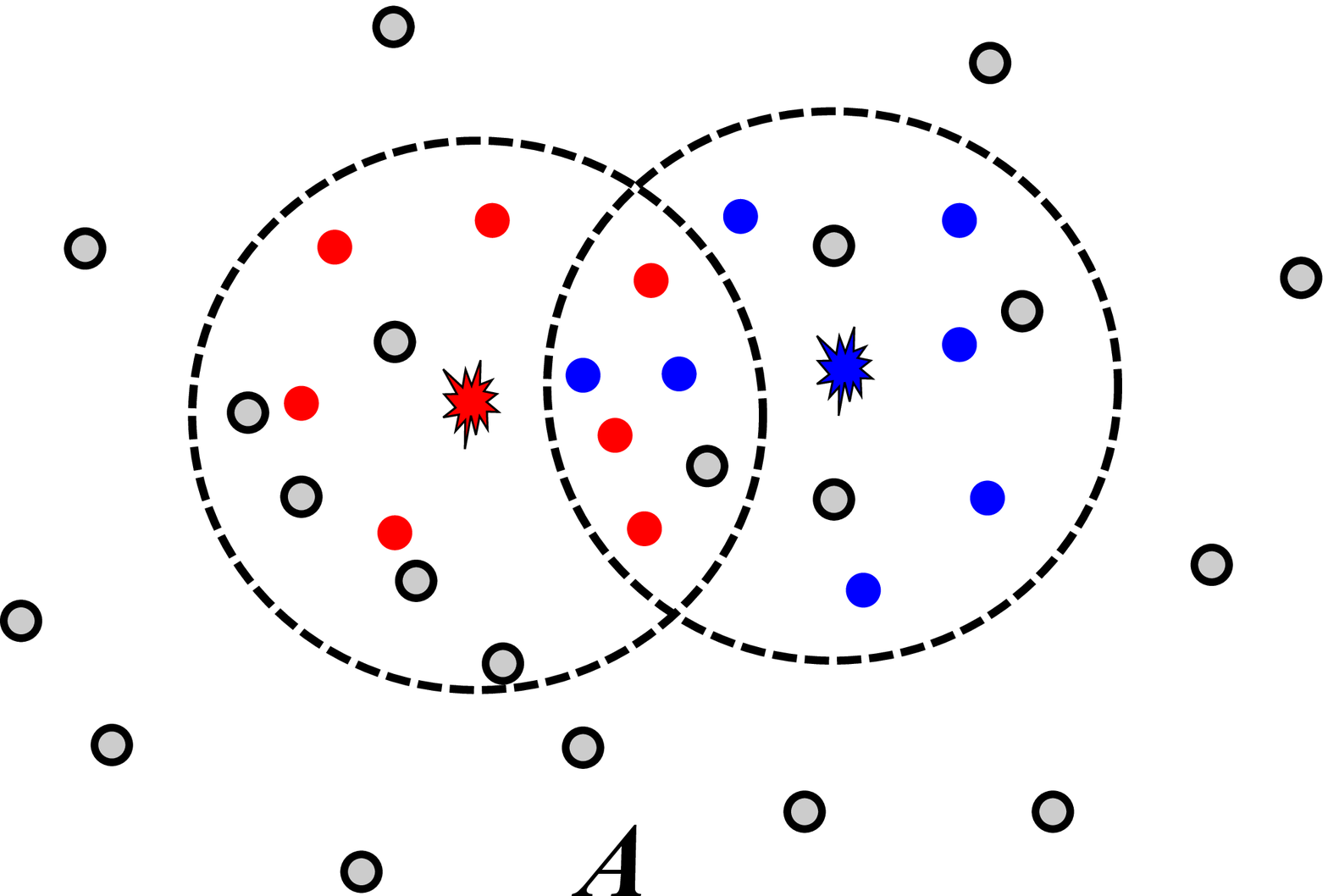}}\hspace{1cm}
   {\includegraphics[width=2.2in,height=1.5in]{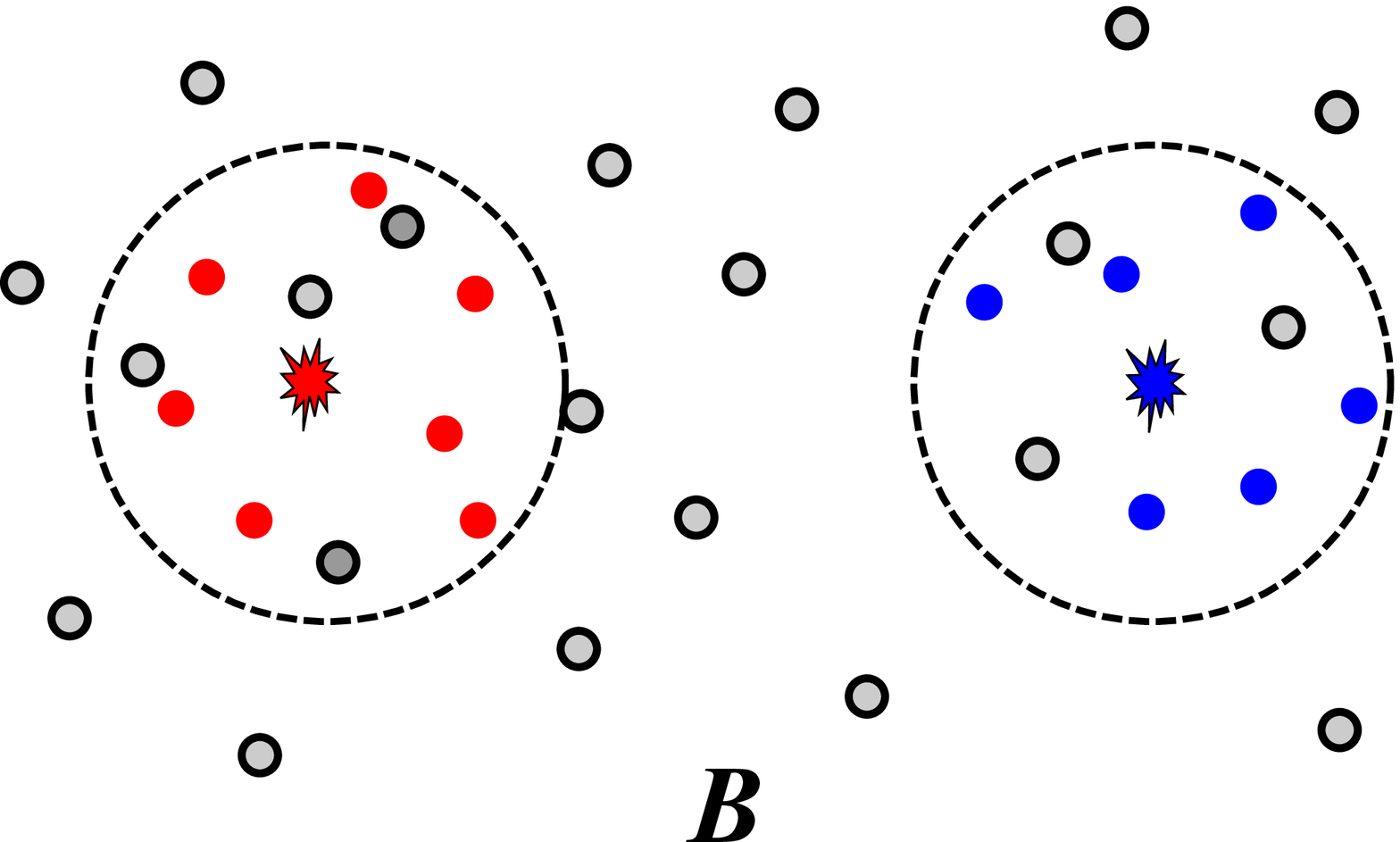}}}
    \caption{Examples of dense (A) and sparse (B) wireless sensor networks.}
    \label{F:F6b}
\end{figure}

\begin{figure}
  \centerline{\includegraphics[width=4.2in]{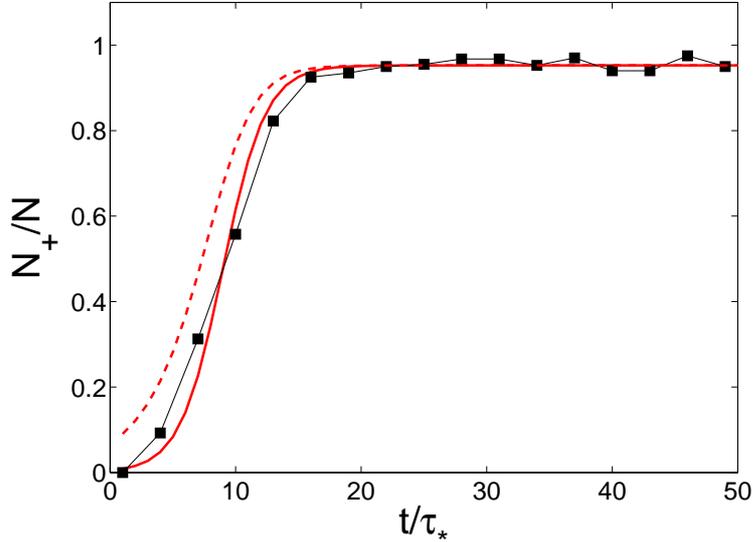}}
    \caption{Effect of parameter $\nu$ in the model (\ref{eqn:E61}) on the simulation data fit: dashed line $\nu = 1$, solid line $\nu=0.7$.}
 \label{F:F7}
\end{figure}

The disagreement described above is due to the implicit assumption
of ``homogeneous mixing'' which we made in equations
(\ref{eqn:E5})-(\ref{eqn:E5a}). The homogeneous mixing manifests
itself in the bilinear form of the interaction terms on the RHS of
(\ref{eqn:E5})-(\ref{eqn:E5a}). This bi-linearity means that the
number of new ``infected'' sensors is proportional to the product of
the number which is currently ``infected'' and the number which is
currently ``susceptible''. Effectively it means that all passive
sensors are equally likely to be activated. This assumption holds
only if the majority of activated (``infected'') sensors are far
away from each other (i.e. at the distances $\gg r_*$). At some
stage of the sensor ``epidemic'' this assumption can be violated,
because the secondary ``infected'' sensors will be  at the shorter
distances from the ``infectious'' parents (see Fig.\ref{F:F6b}). The
broadcasted messages in overlapping areas become duplicated and the
rate of new ``infections''  will be no longer proportional to the
number of their parents. The fraction of ``infected'' sensors in the
overlapping areas will depend on the new equilibrium state of the
sensor system (i.e. $N_+/N$ as $t \rightarrow \infty$) and may not
be small for some scenarios. To overcome this restriction we again
invoke an approach successfully implemented in epidemiology (see
\cite{R10}). Instead of (\ref{eqn:E5})-(\ref{eqn:E5a}) we now write
\begin{equation}\label{eqn:E61}
      \frac{d N_{+}}{dt} =   \alpha N^{\nu}_{+} N_{-}  -
      \frac{N_{+}}{\tau_*}, ~~
      \frac{d N_{-}}{dt} =   - \alpha N^{\nu}_{+} N_{-}  +
      \frac{N_{+}}{\tau_*},
\end{equation}
where a new parameter $0 \leq {\nu} \leq 1$ depends on the packing
density of ``infected'' sensors (or on the ratio $N_+/N$). For a
``sparse'' network configuration we have $\nu \approx 1$ (no
overlapping areas) and for an extremely ``dense'' network $\nu
\approx 0$ (all sensors are located around the same point), see Fig.
\ref{F:F6b}.  In general $\nu$ can be used as a fitting parameter of
the model \cite{R4} or estimated based on the mathematical theory of
packing. For a specific network configuration a value $\nu = 1/2$
was derived in \cite{R5} based on some simplified assumptions. By
employing new parameter $\nu$ we can significantly improve agreement
between analytical model and simulation at the initial stage of
information epidemic, since here we can assume $N_{-} \approx N = const$, so
$\frac{d N_{+}}{dt} \propto  N^{\nu}_{+}$. An example of
improved fitting is presented in Fig.\ref{F:F7}.

Similarly to the epidemiological models (see \cite{R9}),
incorporation of the spatial inhomogeneity can be achieved by adding
the appropriate diffusion terms on the LHS of (\ref{eqn:E5}) and
(\ref{eqn:E5a}):
\begin{equation}\label{eqn:E51}
      \frac{\partial N_{+}}{\partial t} - D \triangle  N_{+} =   \alpha N_{+} N_{-}  -
      \frac{N_{+}}{\tau_*}, ~~
\end{equation}
\begin{equation}\label{eqn:E52}
\frac{ \partial N_{-}}{\partial t} - D \triangle  N_{-}  =   -
\alpha N_{+} N_{-}  +
      \frac{N_{+}}{\tau_*},
\end{equation}
where $D$ is diffusivity in the sensor system which can be estimated
as $D \approx r^{2}_*/\tau_*$. At the same time the inhomogeneity of
pollutant distribution can be easily incorporated in
$\alpha(\textbf{r})$ with non-uniform $C_0(\textbf{r})$ (see
(\ref{eqn:E1}), (\ref{eqn:E4}), (\ref{eqn:E110})).

An important property of the system (\ref{eqn:E51}), (\ref{eqn:E52})
is the existence of analytical solutions in the form of traveling
waves, propagating with the velocity $v_0 \sim \sqrt{\alpha D}$
\cite{R9}. In our case these waves correspond to the switching
fronts between active and passive sensors. If pollutant is advected
by the wind flow with a characteristic velocity $v_*$, then a simple
synchronisation condition $v_0 \geq  v_*$ or $\alpha \geq v^{2}_*
\tau_*/r^{2}_* $ provides an important criteria for network
optimisation.

Another interesting extension of the proposed model is the
introduction of the concept of a {\em faulty} sensor,  a sensor
which is no longer available for sensing and networking. This state
of a sensor would correspond to the {\em removed} population segment
in the epidemiological framework and can be attributed to any kind
of faults (flat battery, software malfunction, hardware defects
etc). As in the celebrated SIR epidemiological model \cite{R9}, a
new state results in the third equation for $N_0$ in the system
(\ref{eqn:E5})-(\ref{eqn:E5a}) with a new temporal parameter - an
average operational time (the lifespan) of a sensor. The total
number of sensors will be still conserved: $N = N_+ + N_- + N_0 =
const$. This model provides a more realistic representation of an
operational sensor systems and allows us to estimate such important
parameters as the operational lifetime of the network and the
reliability of the network.


\section{Conclusions}

We developed a ``bio-inspired'' model of a network of chemical
sensors with dynamic collaboration for the purpose of energy
conservation and information gain. The proposed model leverages on
the existing theoretical discoveries from epidemiology resulting in
a simple analytical model for the analysis of network dynamics. The
analytical  model enabled us to formulate analytically the
conditions for the network performance. Thus we  found an optimal
configuration which, within the underlying assumptions, yields a
balance between the number of sensors, detected concentration, the
sampling time and the communication range. The findings are partly
supported by numerical simulations. Further work is required to
address the model refinements and generalisations.

\section{Acknowledgement}

The authors would like to thank   Ralph
Gailis,  Ajith Gunatilaka and  Chris Woodruff for helpful technical discussions and Champake Mendis
for his assistance in the software
implementation of the proposed model.


\begin{thebibliography}{99}




\bibitem{R0}

C. S. Raghavendra,  K. M. Sivalingam, Taieb Znati. (2005) Wireless
Sensor Networks. Springer, USA, 2005.


\bibitem{R14}
P. E. Bieringer, A. Wyszogrodzki, J. Weil and G. Bieberbach. An Evaluation of Propylene Sampler Grid Designs for the FFT07 Field Program(2006), Tech. Report, National Center for Atmospheric Research, Boulder, USA.





\bibitem{R1}

E. Ertin, J. W. Fisher, L. C. Potter.(2003)  Maximum Mutual
Information Principle for Dynamic Sensor Query Problems.
\emph{Lecture Notes in Computer Science: Information Processing in
Sensor Networks},  2003 \textbf{2634}, pp. 91-104.

\bibitem{R2}

F.  Zhao, J. Shin, J. Reich. (2002) Information-Driven Dynamic
Sensor Collaboration for Tracking Applications. \emph{IEEE Signal
Processing Magazine}, 2002, \textbf{19}, 2, pp. 61--72.

\bibitem{R22}

J. Mathieu, G. Hwang, J. Dunyak (2006). The State of the Art and the
State of the Practice: Transferring Insights from Complex Biological
Systems to the Exploitation of Netted Sensors in Command and Control
Enterprises, \emph{2006 MITRE Technichal Papers}, July, 2006, MITRE
Corporation, USA.


\bibitem{R3}

A. Khelil, C. Becker, J. Tian, K. Rothermel.(2002) An Epidemic Model
for Information Diffusion in MANETs.In \emph{MSWiM 2002: Proceedings
of the 5th ACM international workshop on Modeling analysis and
simulation of wireless and mobile systems}, Atlanta, Georgia, USA,
2002, pp. 54--60.


\bibitem{R5}


P. De, Y. Liu, S. K. Das. (2007) An Epidemic Theoretic Framework for
Evaluating Broadcast Protocols in Wireless Sensor Networks. In
\emph{MASS 2007: Proceedings of IEEE Internatonal Conference on
Mobile Adhoc and Sensor Systems},   Pisa, Italy, 2007, pp 1--9.

\bibitem{R00}

A. Gunatilaka, B.Ristic, A.Skvortsov, M. Morelande. (2008) Parameter
Estimation of a Continuous Chemical Plume Source. in \emph{Fusion
2008: 11th International Conference on Information Fusion}, Cologne,
Germany, 2008, pp. 1-8.


\bibitem{R4}

B.Ristic, A.Skvortsov, M.Morelande. (2009) Predicting the Progress
and the Peak of an Epidemics. in \emph{ICASSP 2009: Proceedings of
2009 IEEE International Conference on Acoustics, Speech and Signal
Processing}, Taiwan, April, 2009.



\bibitem{R6}

A.Dekker, A.Skvortsov. (2009)  Topological Issues in Sensor
Networks.  in \emph{MODSIM 2009: 2009 MSSANZ International Congress
on Modelling and Simulation}, Cairns, Australia, 2009.


\bibitem{R7}

S. Eubank, V. S. Anil Kumar, M. Marathe. (2008)  Epidemiology and
Wireless Communication: Tight Analogy or Loose Metaphor?
\emph{Lecture Notes in Computer Science: Bio-Inspired Computing and
Communication}, 2008, \textbf{5151}, pp. 91-104.


\bibitem{R9}

J. D. Murray.(2002) Mathematical Biology, Springer, USA, v 1,2. 2002


\bibitem{bisiganesi_07}
V. Bisignanesi, M.S. Borgas (2007). Models for integrated pest
management with chemicals in atmospheric surface layers. \emph{{
Ecological modelling}},  2007, \textbf{201}, 1, pp. 2--10.



\bibitem{R8}

P. D. Stroud, S. J. Sydoriak, J. M. Riese, J. P. Smith, S. M.
Mniszewski, and P. R. Romero. (2006) Semi-empirical Power-law
Scaling of New Infection Rate to Model Epidemic dynamics with
Inhomogeneous Mixing, \emph{Mathematical Biosciences}, \textbf{203},
pp. 301--318.



\bibitem{R10}

A.T. Skvortsov,  R.B.Connell, P.D. Dawson, R.M. Gailis. (2007)
Epidemic Spread Modeling: Alignment of Agent-Based Simulation with a
Simple Mathematical Model. in \emph{BIOCOMP 2007: Proceedings of
International Conference on Bioinformatics and Computational
Biology}, Las Vegas Nevada, USA, CSREA Press, 2007, \textbf{2}, pp
487--490.

\bibitem{R11}

A. Gunatilaka, A. Skvortsov, and R. Gailis. (2008) Progress in DSTO CBR simulation environment development," Land Warfare Conference (LWC2008), Brisbane, 2008, pp. 62--68.

\bibitem{R12}

A.Skvortsov, E.Yee. Scaling laws of peripheral mixing of passive scalar in a wall-shear layer (2008). \emph{ Phys. Rev. E} \textbf{83}, 036303--11.

\bibitem{R13}
M. Jamriska, T. C. DuBois, A. Skvortsov Statistical characterisation of bio-aerosol background in an urban environment (2011),
http://www.arxiv.org/pdf/1110.4184


\end{thebibliography}
\end{document}